\documentclass[twocolumn,showpacs,preprintnumbers,amsmath,amssymb,superscriptaddress,nofootinbib]{revtex4-1}

\usepackage{graphicx}
\usepackage{dcolumn}
\usepackage{bm}

\begin{document}

\title{Measurement of the $^{70}$Ge($n,\gamma$) cross section up to 300~keV at the CERN n\_TOF facility}

\author{A.~Gawlik} \affiliation{University of Lodz, Poland} %
\author{C.~Lederer-Woods} 
\email[corresponding author: ]{claudia.lederer-woods@ed.ac.uk}
\affiliation{School of Physics and Astronomy, University of Edinburgh, United Kingdom} %
\author{J.~Andrzejewski} \affiliation{University of Lodz, Poland} %
\author{U.~Battino} \affiliation{School of Physics and Astronomy, University of Edinburgh, United Kingdom} %
\author{P.~Ferreira} \affiliation{Instituto Superior T\'{e}cnico, Lisbon, Portugal} %
\author{F.~Gunsing} \affiliation{CEA Irfu, Universit\'{e} Paris-Saclay, F-91191 Gif-sur-Yvette, France} \affiliation{European Organization for Nuclear Research (CERN), Switzerland} %
\author{S.~Heinitz} \affiliation{Paul Scherrer Institut (PSI), Villingen, Switzerland} %
\author{M.~Krti\v{c}ka} \affiliation{Charles University, Prague, Czech Republic} %
\author{C.~Massimi} \affiliation{Istituto Nazionale di Fisica Nucleare, Sezione di Bologna, Italy} \affiliation{Dipartimento di Fisica e Astronomia, Universit\`{a} di Bologna, Italy} %
\author{F.~Mingrone} \affiliation{Istituto Nazionale di Fisica Nucleare, Sezione di Bologna, Italy} %
% ...
%

\collaboration{The n\_TOF Collaboration (www.cern.ch/ntof)} \noaffiliation

\date{\today}

\begin{abstract}
Neutron capture data on intermediate mass nuclei are of key importance to nucleosynthesis in the weak component of the slow neutron capture processes, which occurs in massive stars.
The ($n,\gamma$) cross section on $^{70}$Ge, which is mainly produced in the $s$~process, was measured at the neutron time-of-flight facility n\_TOF at CERN. 
Resonance capture kernels were determined up to 40~keV neutron energy, and average cross sections up to 300~keV. 
Stellar cross sections were calculated from $kT=5$~keV to $kT=100$ keV and are in very good agreement with a previous measurement by Walter and Beer (1985),
and recent evaluations. Average cross sections are in agreement with Walter and Beer (1985) over most of the neutron energy range covered, while being systematically smaller for neutron energies above 150~keV. 
We have calculated isotopic abundances produced  in $s$-process environments in a 25 solar mass star for two initial metallicities (below solar, and close to solar). While the low metallicity model reproduces best the solar system germanium isotopic abundances the close to solar model shows a good global match to solar system abundances between mass numbers A=60-80. 

\end{abstract}


\maketitle

\section{\label{intro}Motivation}
\begin{figure}[!htb]
\includegraphics[width=0.4\textwidth]{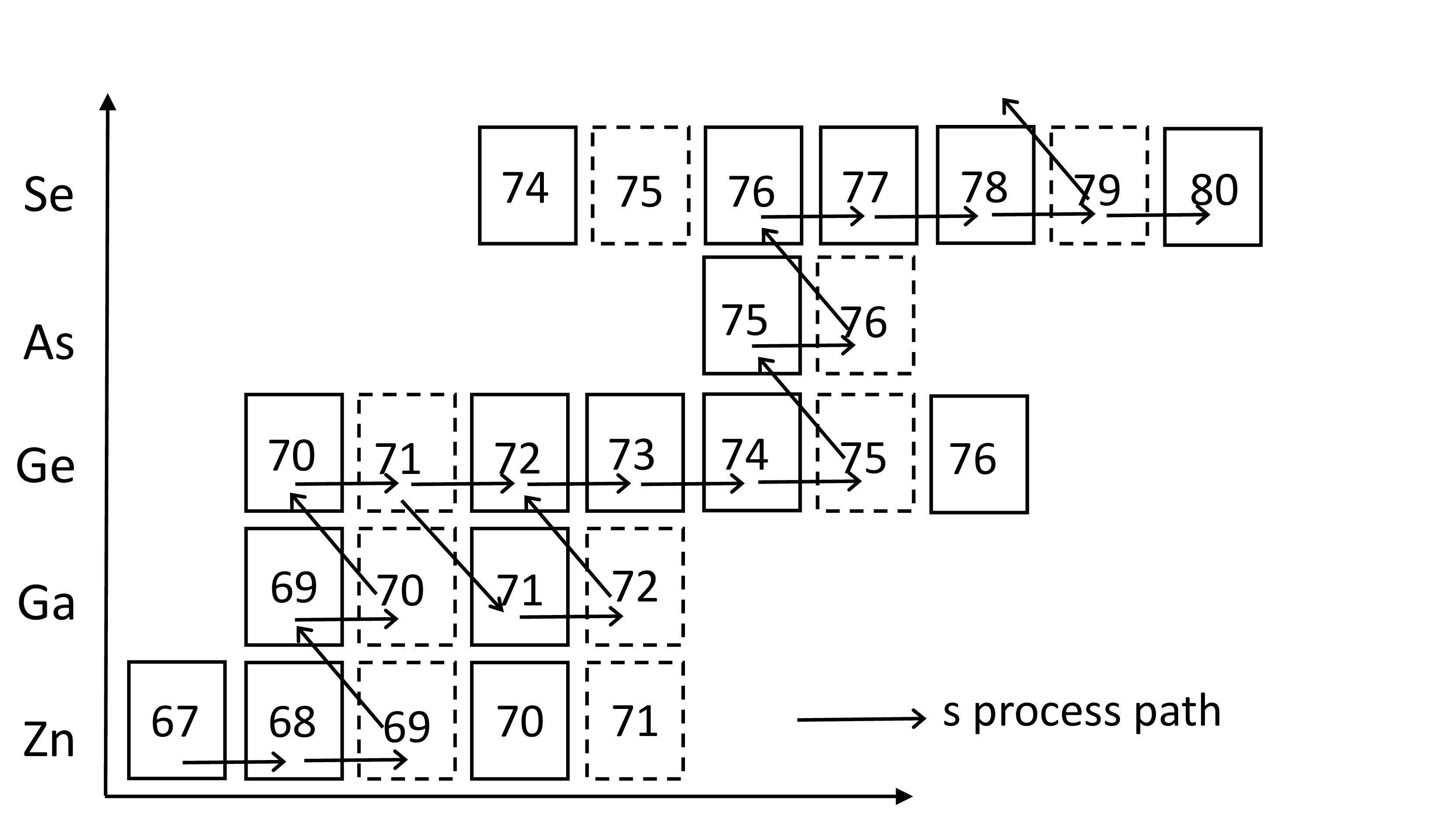}
\caption{Main nucleosynthesis path of the $s$ process in massive stars during He core burning. Solid boxes represent stable, dashed boxes represent unstable isotopes. \label{spath}}
\end{figure}

The elemental abundances above Fe are mainly produced by two distinct neutron capture processes in stars and stellar explosions, the slow neutron capture process ($s$~process) \cite{RLK14} and the rapid neutron capture process ($r$~process) \cite{TAK11}, while only about 1\% of heavy element abundances is produced by charged particle and photo-induced reactions ($p$~process) \cite{RDD13}.
The $s$~process occurs at moderate neutron densities of typically $10^{7}-10^{12}$ cm$^{-3}$, and is characterised by a sequence of neutron captures followed by $\beta$-decays \cite{RLK14}.
Since radioactive decays are usually faster than neutron capture times, the reaction path proceeds along the line of stable nuclei. 
In massive stars ($> 8$ solar masses $M_{\odot}$), the $s$~process occurs in two different burning stages, towards the end of He core burning at temperatures of about 0.3 GK (1 GK=$10^9$~K), 
and later during carbon shell burning, when temperatures reach 1 GK \cite{PET68,COUCH74,LAMB77,RAIT91a,RAIT91b}. 
This so called $weak$ component of the $s$~process, produces dominantly elements between Fe and Zr, owed to the relatively low neutron exposures reached. The $main$ component of the $s$~process
contributes dominantly abundances between Zr and Bi, and occurs in low and intermediate mass stars  (about 1-5 $M_{\odot}$) during their Asymptotic Giant Branch (AGB) phase \cite{RLK14, H05}.\\
In our cosmos, germanium is produced by more than one nucleosynthesis process, however, the bulk of it (about 80\%) is thought to be produced by the weak $s$-process in massive stars \cite{PIG10}. 
Around 12\% are estimated to originate from the $main$ $s$-process in AGB stars, while only a few percent are produced by explosive nucleosynthesis processes, such as 
the $r$ process, operating at very high neutron densities, and the alpha-rich freeze out forming the Fe group nuclei at the end of a massive star's life. 
To disentangle the different contributions to observed germanium abundances in our solar system and stellar atmospheres, it is crucial to accurately determine the $s$ process contribution to germanium abundances.
Figure \ref{spath} shows the $s$-process reaction path in the germanium mass region. $^{70}$Ge occupies a special position: It is shielded from
rapid neutron capture nucleosynthesis by the stable isobar $^{70}$Zn, suggesting that $^{70}$Ge is dominantly produced in the $s$~process (so-called s-only isotope), and its abundance thus can be used as an anchor point when 
comparing to observed isotopic abundances. \\
Neutron capture cross sections are a key nuclear physics input to calculate abundances produced in $s$-process nucleosynthesis. Since neutrons are rapidly thermalised in the stellar environment,
the effective stellar cross section is a convolution of the energy dependent neutron capture cross section with a Maxwellian velocity distribution, thus, stellar neutron capture cross sections 
are called Maxwellian Averaged Cross Sections (MACS). For $s$-process environments, MACS values need to be known up to $kT$ values of about 90~keV, corresponding to the highest temperatures reached in the 
$s$ process. \\
Investigations using Monte Carlo variation of reaction rates to assess
the impact of specific reactions on the final abundances predicted by
stellar models have been performed recently. The reaction $^{70}$Ge($n,\gamma$) has
been identified as a key reaction determining the uncertainty of $^{70}$Ge in
the main s-process, both in the thermal pulse and in interpulse burning
of AGB stars \cite{cesc18}. The reaction $^{70}$Ge($n,\gamma$) and the recently measured
$^{73}$Ge($n,\gamma$) \cite{LB19} were found to be key reactions for the abundances of
$^{70}$Ge and $^{73}$Ge, respectively, in the enhanced weak s-process occurring in
rotating, metal-poor massive stars \cite{nishi17}. $^{73}$Ge($n,\gamma$) also appeared as
key rate in the regular weak s-process in the same study. \\
Existing experimental data on stellar neutron capture on germanium isotopes are scarce, which motivated a 
campaign at n\_TOF to measure cross sections on all stable germanium isotopes \cite{Gepro}. Results on 
$^{73}$Ge($n,\gamma$) have recently been published and show that the germanium isotopic abundance pattern produced in the $s$~process in a low metallicity massive star is 
consistent with the isotopic abundance pattern of our solar system \cite{LB19}. This manuscript describes the study of the $^{70}$Ge($n,\gamma$) reaction.
There are few experimental data available on this reaction in the keV neutron energy range; 
Walter and Beer \cite{WB85} measured cross sections from 3 to 240~keV and calculated  MACSs between $kT=20-50$~keV. In addition,
Maletski et al. \cite{Mal68} have published neutron resonance data on $^{70}$Ge$+n$ reactions for neutron energies up to 28.6 keV, however, partial radiative widths $\Gamma_\gamma$
have only been determined for three low energy resonances, while for the remaining resonances only neutron widths $\Gamma_n$ and resonance energies $E_R$ are available. 
Harvey and Hockaday \cite{HH80} have determined total cross sections for natural germanium, identifying several $^{70}$Ge$+n$ resonances up to 43 keV neutron energy. 
Evaluated $^{70}$Ge($n,\gamma$) cross sections recommended in major nuclear data libraries such as ENDF/B-VIII \cite{ENDF8} and JENDL-4.0 \cite{JENDL40} use 
experimental data by Maletski at al. and transmission data on natural germanium by Harvey and Hockaday for neutron energies below 14~keV, 
while for higher neutron energies evaluated cross sections are based on statistical parameters extracted from data at lower neutron energies. \\
This work provides neutron capture resonance data up to energies of 40~keV. In combination with average neutron capture cross sections determined for neutron energies up to
300~keV, we also calculated Maxwellian averaged cross sections for the entire range of astrophysical interest, from $kT=5$~keV, to $kT=100$~keV.

\section{\label{measurement}Measurement}
The measurement was performed at the neutron time-of-flight facility n\_TOF at CERN \cite{ntofpage}. At n\_TOF, neutrons are produced by spallation reactions of a 20 GeV/c proton beam
by the CERN-PS on a cylindrical 1.3 tonne lead target (40~cm length, 60~cm diameter) \cite{guerrero2012}. 
The initially highly energetic neutrons are  moderated using a combination of water and borated water layers surrounding the target (the former also cools the target). 
The resulting moderated neutron spectrum exhibits a nearly isolethargic energy dependence and ranges from thermal energies (25 meV) to several GeV. 
The proton beam is pulsed, with a time width of 7 ns r.m.s., and a repetition rate of about 0.8 Hz. At n\_TOF, there are two experimental areas at flight path lengths of 185~m (EAR-1) and 20~m (EAR-2). The $^{70}$Ge($n,\gamma$) measurement was
performed at EAR-1, taking advantage of the excellent neutron energy resolution ranging from  $3\times10^{-4}$ at 1 eV to $3\times10^{-3}$ at 100 keV \cite{guerrero2012}. \\
The prompt emission of $\gamma$ rays following neutron capture was detected with a set of four liquid scintillation detectors, filled with deuterated benzene (C$_6$D$_6$). These detectors
are optimal for measurement of radiative neutron capture since they are almost insensitive to neutrons which are scattered from the capture sample. The detectors were arranged at angles of 125
degrees with respect to the neutron beam to minimise effects of anisotropic $\gamma$-ray emission for $\ell>0$ states in the compound nucleus. \\
The $^{70}$Ge($n,\gamma$) reaction was studied using a highly enriched $^{70}$GeO$_{2}$ sample in cylindrical form.  
In addition, measurements were taken with a metallic germanium sample of natural composition to identify any contributions from other Ge isotopes present in the enriched sample, and with a Au sample
which was used to normalise the neutron capture data (see Sec. \ref{analysis}). The samples were glued on a thin mylar foil fixed to an Al ring. The background
induced by the sample holder was measured by placing an empty sample holder in the beam. 
To keep any systematic effects due to sample positioning to a minimum, the samples were accurately centred on the holder using a jig and a hollow metallic cylinder aligned with the annular frame of the sample holder. 
Table \ref{samtab} lists all samples and their characteristics used in the measurement. 

\begin{table*}[!htb]
\caption{Properties of the samples used in the experiment \label{samtab}
}
\begin{ruledtabular}
\begin{tabular}{ccccc}

Sample & Chemical Form & Mass (g) & Diameter (cm) & Sample Composition (\%) \\ \hline
$^{70}$Ge & GeO$_2$ & 2.705 & 2 & $^{70}$Ge(97.71); $^{72}$Ge(2.23); $^{73}$Ge(0.02); $^{74}$Ge(0.03); $^{76}$Ge(0.01)  \\
$^{\text{nat}}$Ge & metal & 1.903 & 2 & $^{70}$Ge(20.52); $^{72}$Ge(27.45); $^{73}$Ge(7.76); $^{74}$Ge(36.52); $^{76}$Ge(7.75) \\
$^{197}$Au & metal & 0.664 & 2 & $^{197}$Au (100) \\
Empty Holder & - & - & - & - \\
\end{tabular}

\end{ruledtabular}
\end{table*}

Detector signals were recorded using 14-bit fast digitizers operated at a sampling rate of 1~GHz. Data acquisition was triggered by the 
pickup signal of the PS-accelerator and for each neutron pulse data were recorded for a duration of 100 ms corresponding to neutron energies of about 0.02~eV. Detector signal times and amplitudes were extracted for each neutron burst using an off-line 
pulse shape algorithm \cite{Zugec16}.

\section{\label{analysis}Data Analysis and Results}
\subsection{Time-of-flight to neutron energy conversion}

The neutron time-of-flight spectra were converted to neutron energy using the relativistic relation
\begin{equation}
 E_n=m_n c^2(\gamma-1)
\end{equation}
with 
\begin{equation}
 \gamma=\frac{1}{\sqrt{1-(L/t_f)^2/c^2}}
\end{equation}
where $m_n$ is the mass of the neutron, $L$ is the flight path length, and $t_f$ is the neutron time-of-flight. The determination of $t_f$ relies on an accurate knowledge 
of the neutron production time. When the proton pulse hits the spallation target, a burst of $\gamma$-rays and ultra-relativistic particles is produced alongside neutrons. 
This burst is called $\gamma$-flash, and produces a short, high amplitude signal in the C$_6$D$_6$ detectors, allowing determination of the time of neutron production for each pulse with high accuracy. 
The flight path length $L$ was determined using a Au sample in the beam. Au has several low energy resonances for which resonance energies have been accurately determined at 
the Joint Research Centre in Geel \cite{MBK11}. These are now included in the latest
evaluated resonance data for $^{197}$Au($n,\gamma$) by ENDF/B-VIII \cite{ENDF8}.
The flight path has been fitted to reproduce the resonance energies for these well known resonances and determined as $183.94\pm0.04$~m. 

\subsection{Capture Yield}
The neutron capture yield as a function of neutron energy is obtained as 
\begin{equation}
 Y(E_n)=f_N(E_n) \frac{C(E_n)-B(E_n)}{\epsilon_c \Phi(E_n)}
\end{equation}
where $C$ is the count spectrum of the enriched $^{70}$Ge sample, $B$ is the background, $\epsilon_c$ is the efficiency to detect a capture event, and $\Phi$ is the neutron flux. 
The factor $f_N$ is a normalisation factor taking into account that the capture sample does not cover the entire size of the neutron beam. In the following sub-sections, 
the determination of each of those terms will
be described. 

\subsubsection{Detection efficiency}
The efficiency to detect the $\gamma$-ray emission after a neutron capture event depends on the de-excitation path of the compound nucleus, which in general varies from event to event. The total energy detection principle was used by combining the above-mentioned detection system with the Pulse Height Weighting Technique (PHWT) \cite{pwht1,pwht3}. 
The total energy detection principle is applicable to detection systems of low efficiency, where at most one $\gamma$ ray per capture event is detected. 
If the $\gamma$-ray detection efficiency is proportional to the $\gamma$-ray energy, i.e. $\epsilon_\gamma\propto E_\gamma$,
it can be shown that the efficiency to detect a capture event is proportional to the excitation energy of the compound system, i.e. $\epsilon_c\propto S_n+E_{cm}$, where $S_n$ is the neutron separation
energy, and $E_{cm}$ is the center-of-mass energy. 
To achieve $\epsilon_\gamma\propto E_\gamma$, the PHWT is used, where weighting factors are applied to each detected event, depending on their amplitude. These were determined in GEANT4 Monte Carlo simulations \cite{geant4}, simulating the detector response to  mono-energetic 
$\gamma$-rays from 0.2 to 10.0~MeV. Simulations took into account the detailed geometry of the experimental setup and samples used, including the spatial distribution of the neutron beam,
and neutron and $\gamma$-ray self absorption in the samples.
The data further need to be corrected for a loss of $\gamma$-rays below the detector threshold (200~keV), and for transitions without $\gamma$-ray emission (electron conversion). 
The correction for both Au and Ge  was calculated using the code {\sc dicebox} \cite{Bec98}, that simulates individual capture cascades based on experimental information on low-lying levels and uses level density
and photon strength function models at higher energies. \\
The systematic uncertainty of the neutron capture yield due to the PHWT is 2\% \cite{pwht2}.

\subsubsection{Background Subtraction \label{bgsub}}

\begin{figure}[!htb]
\includegraphics[width=7.0 cm,angle=270]{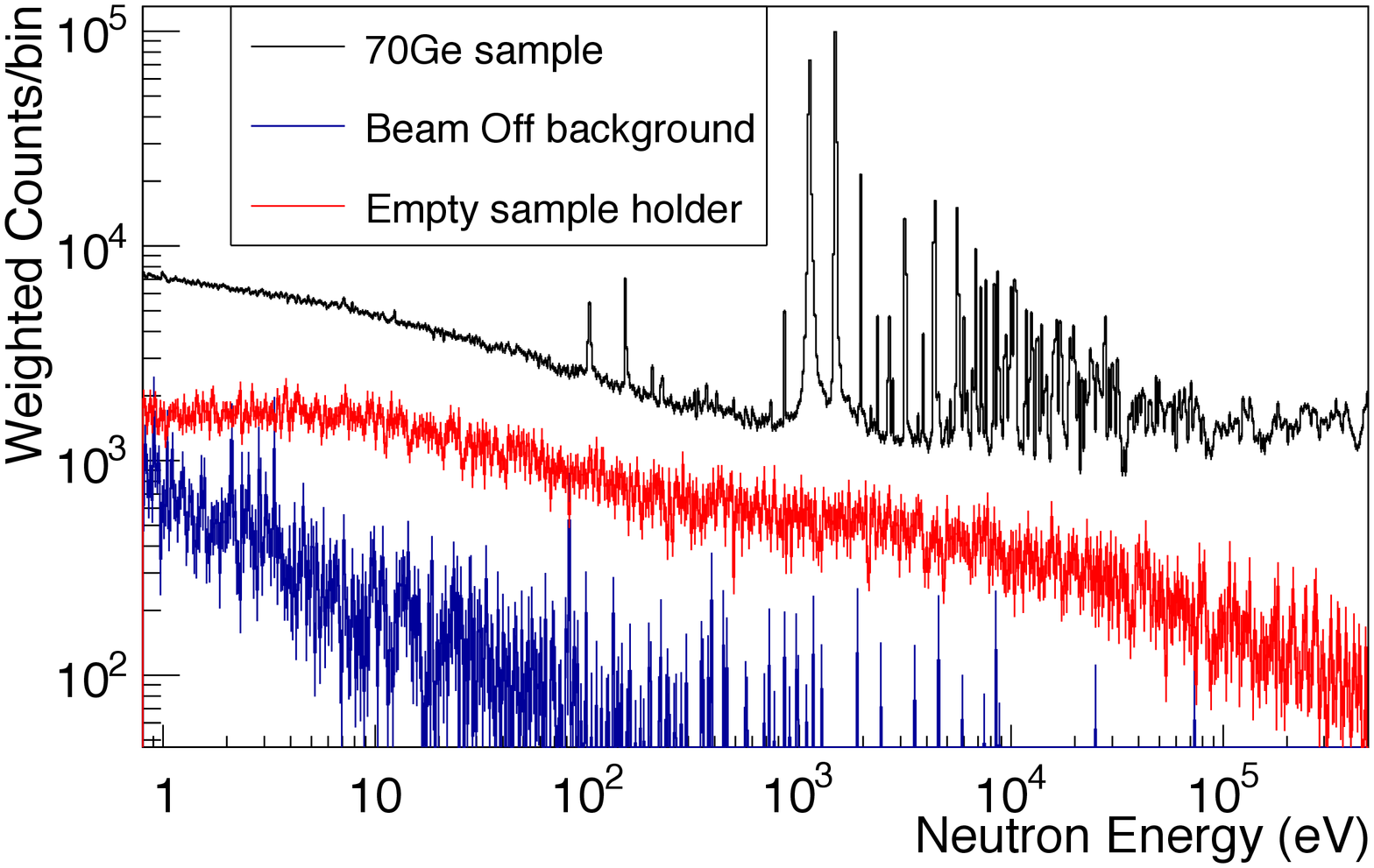}
\caption{(Color online) Plot of the weighted Ge spectrum compared to empty sample holder and ambient background \label{indbg}}
\end{figure}

\begin{figure}[!htb]
\includegraphics[width=7.0 cm,angle=270]{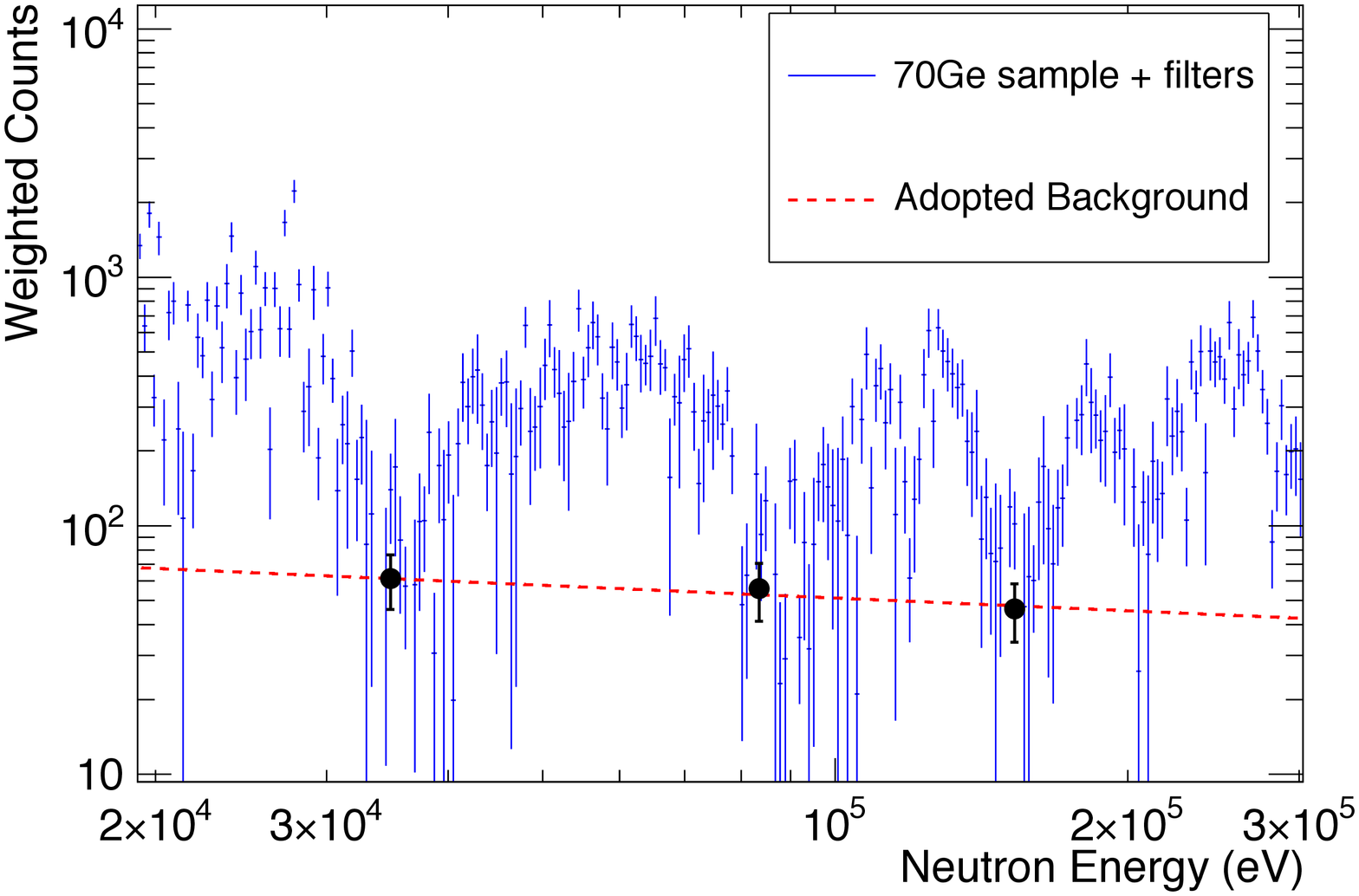}
\caption{(Color online) Ge+filter spectrum (ambient and empty sample holder+filter spectrum subtracted), and background determined from filter dip minima.  \label{filbg}}
\end{figure}

The background $B(E_n)$ consists of several components that need to be corrected for. Background unrelated to the neutron beam, due to ambient 
radioactivity and cosmic rays is determined in beam off runs and subtracted
from the Au and Ge sample spectra. Background produced by neutron reactions with material in and around the beam line is measured by taking data with an empty sample holder. \\
Figure \ref{indbg} shows the $^{70}$Ge spectrum, compared to backgrounds due to the empty sample holder, and due to ambient activity. Weighting functions have been applied to all these spectra, 
hence the Figure shows weighted counts as a function of neutron energy. 
The empty sample holder contributes to background over the entire neutron energy range, while the ambient activity, which is constant in time, only contributes at low neutron energies. 
The third component is background related to the sample, specifically by neutrons scattered off the GeO$_2$ sample. 
These scattered neutrons may be captured in surrounding material and produce $\gamma$-ray cascades detected by the C$_6$D$_6$ system. The C$_6$D$_6$ detectors themselves are extremely insensitive to capturing neutrons \cite{plag, mastinu}, and therefore any background due to direct interaction of scattered neutrons with the detectors themselves can be neglected for the present measurement. 
While care has been taken to minimize material around the beam pipe, and using material with low neutron cross sections, such as Al, there may be residual background due to neutrons scattered from the sample and
being captured at a later time in structural material, such as the walls \cite{Zugec14}. \\
For the resonance region, individual neutron resonances were fitted including a flat, constant background to account for these backgrounds. For the unresolved resonance region, the background is determined using neutron filters. Neutron filters are materials in the beam line, which absorb all neutrons at certain 
resonance energies. In the present case, an Al filter was used, with negligible transmission of neutrons at around 35, 86 and 150 keV neutron energy. Consequently, any counts in these filter dips
must come from neutrons scattered off the sample and captured somewhere in the experimental area at a later time. This background was determined by measuring a Ge spectrum with filters, and an empty sample 
holder spectrum with filters. After subtraction of the empty spectrum,  the Ge+ Al filter spectrum was scaled to the Ge spectrum to account for the overall loss of neutrons. 
A smooth function was fitted to the flat minima of the filter dips and the resulting background was subtracted from the Ge spectrum. Figure \ref{filbg} shows the $^{70}$Ge+filter spectrum and the fitted background. 
The background was found to be at a level of 10-15\% of the Ge spectrum. Due to uncertainties in the scaling factor applied to the Ge+filter spectrum, and low statistics
in the filter dips, the resulting uncertainty of the capture yield in the unresolved resonance region is 4\%. 

\subsubsection{Neutron Flux and Normalisation}
The neutron flux was measured using several different detection systems and reference reactions in a dedicated campaign.
Measurements were performed with a set of silicon detectors, detecting tritons and alpha particles emitted in the $^{6}$Li($n,t$) reaction from a LiF sample. 
A Micromegas detector was used to determine the neutron flux using the reference reactions $^{6}$Li($n,t$) and $^{235}$U($n,f$), and an ionisation chamber by Physikalisch Technische Bundesanstalt Braunschweig 
measured  ($n,f$) reactions on $^{235}$U. The data from all three detection systems were combined to determine an accurate neutron flux, which has a systematic uncertainty below 1\% for neutron energies $<3$~keV, and of 3.5\% between 3~keV 
and 1~MeV \cite{BS16}. More details on the neutron flux evaluation at n\_TOF can be found in Ref. \cite{Barb13}. \\
A normalisation factor $f_N$ needs to be applied since the neutron beam is larger than the capture sample. 
In addition, such a factor corrects for any inaccuracies in determination of the solid angle covered by the 
C$_6$D$_6$ detectors. The saturated resonance technique \cite{mack79} was applied to determine the normalisation factor with high accuracy. In the present case, data were normalised to the 
well known $4.9$-eV resonance in the $^{197}$Au($n,\gamma$) reaction. For this resonance, the capture cross section is much larger than neutron scattering ($\Gamma_\gamma/\Gamma_n\approx8$ \cite{ENDF8}). 
If the Au sample is chosen to be sufficiently thick, all neutrons at the resonance energy will be absorbed by the sample. In addition, due to the high ratio of capture/scattering almost all neutrons will be radiatively captured within the sample, even if their first interaction is elastic scattering. Thus, almost 100\% of neutrons passing the capture sample produce a Au neutron capture cascade, providing an absolute 
normalisation point for the neutron capture yield at 4.9 eV neutron energy. The size of the neutron beam has a slight dependence on neutron energy. This dependence has been determined in Monte 
Carlo simulations of the neutron transport from the spallation target to the experimental area, and has been verified in beam profile measurements \cite{guerrero2012}. In the region of interest, 
the change in neutron beam size is at most 1.5\%. 
This small correction was applied to the data. 
 As the saturated resonance technique is insensitive to the precise individual resonance parameters for the 4.9-eV resonance, the systematic uncertainty, including possible small errors in sample positioning, 
is 1\%.   \\

\subsection{Resonance Analysis \label{RRRsec}}
\begin{figure}[!htb]
\includegraphics[width=7.5 cm]{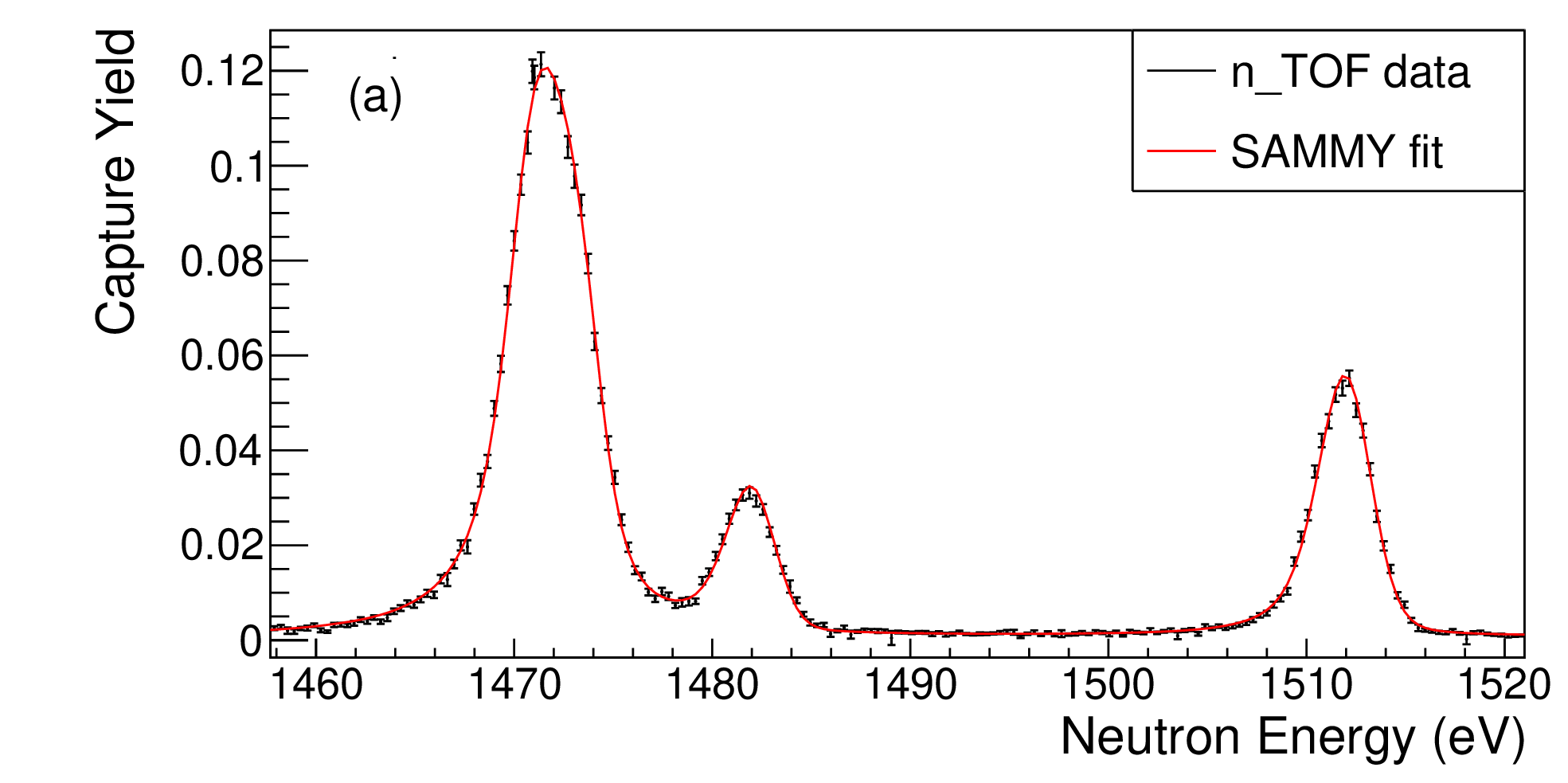}
\includegraphics[width=7.5 cm]{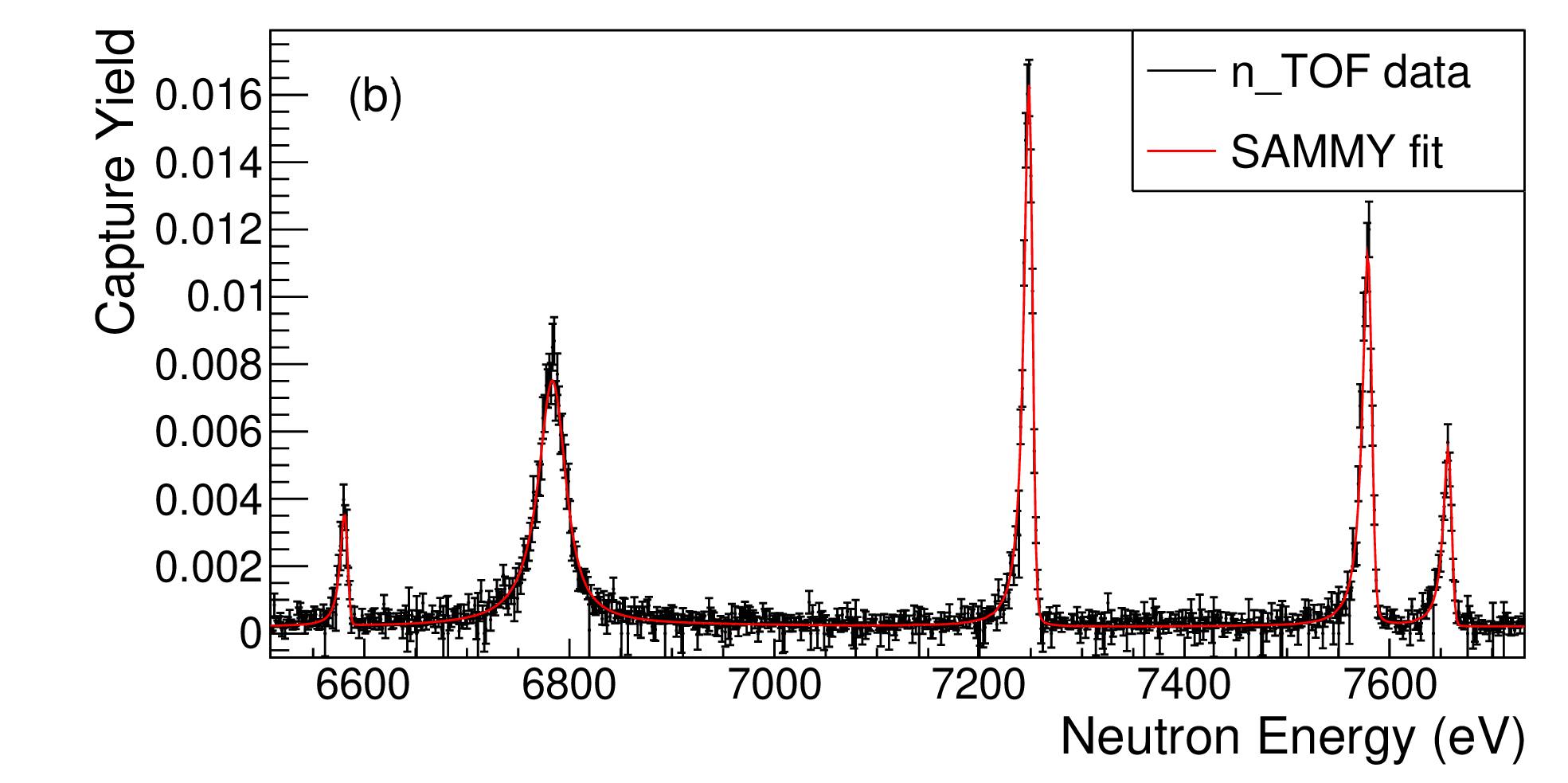}
\includegraphics[width=7.5 cm]{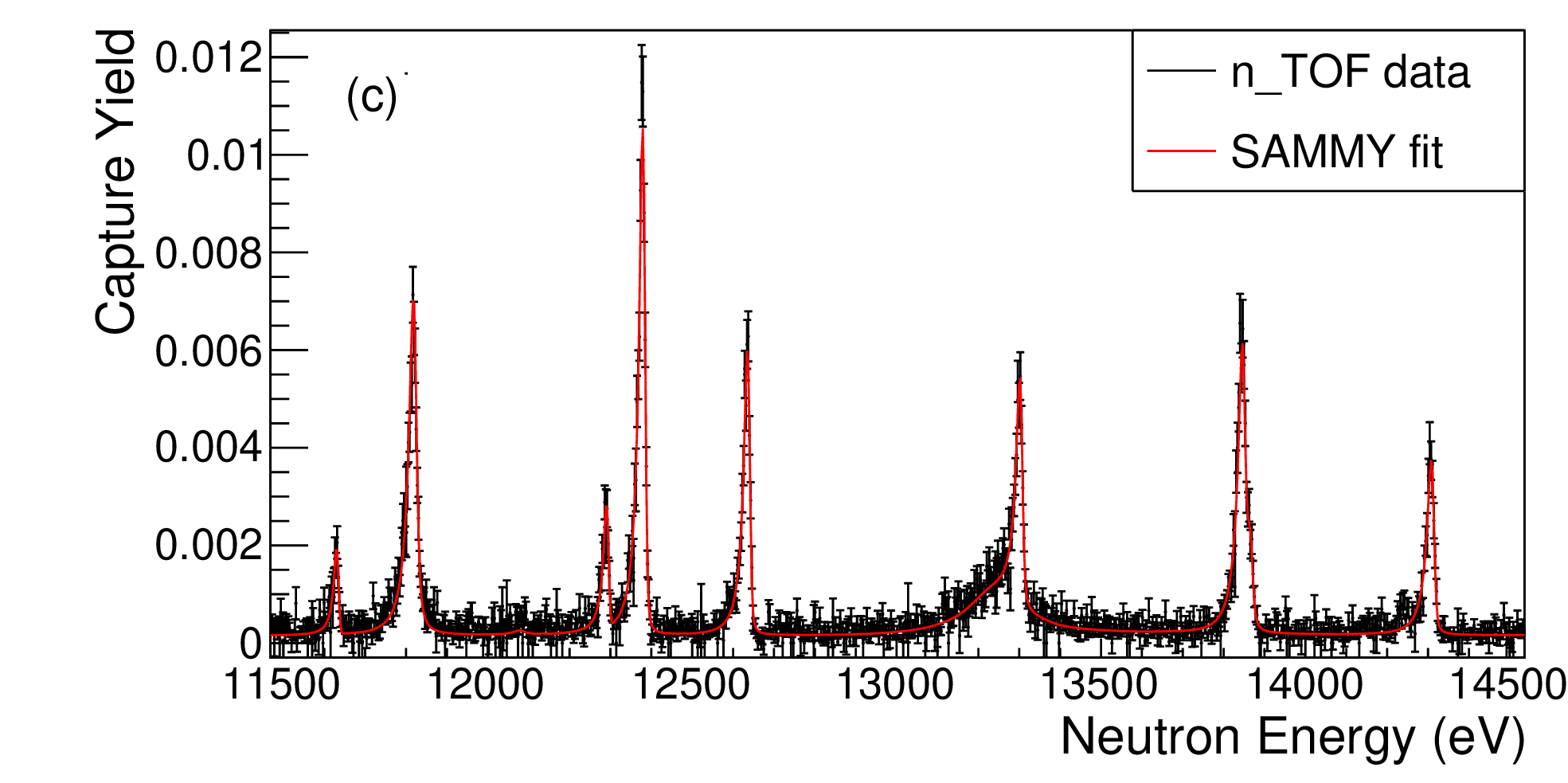}
\includegraphics[width=7.5 cm]{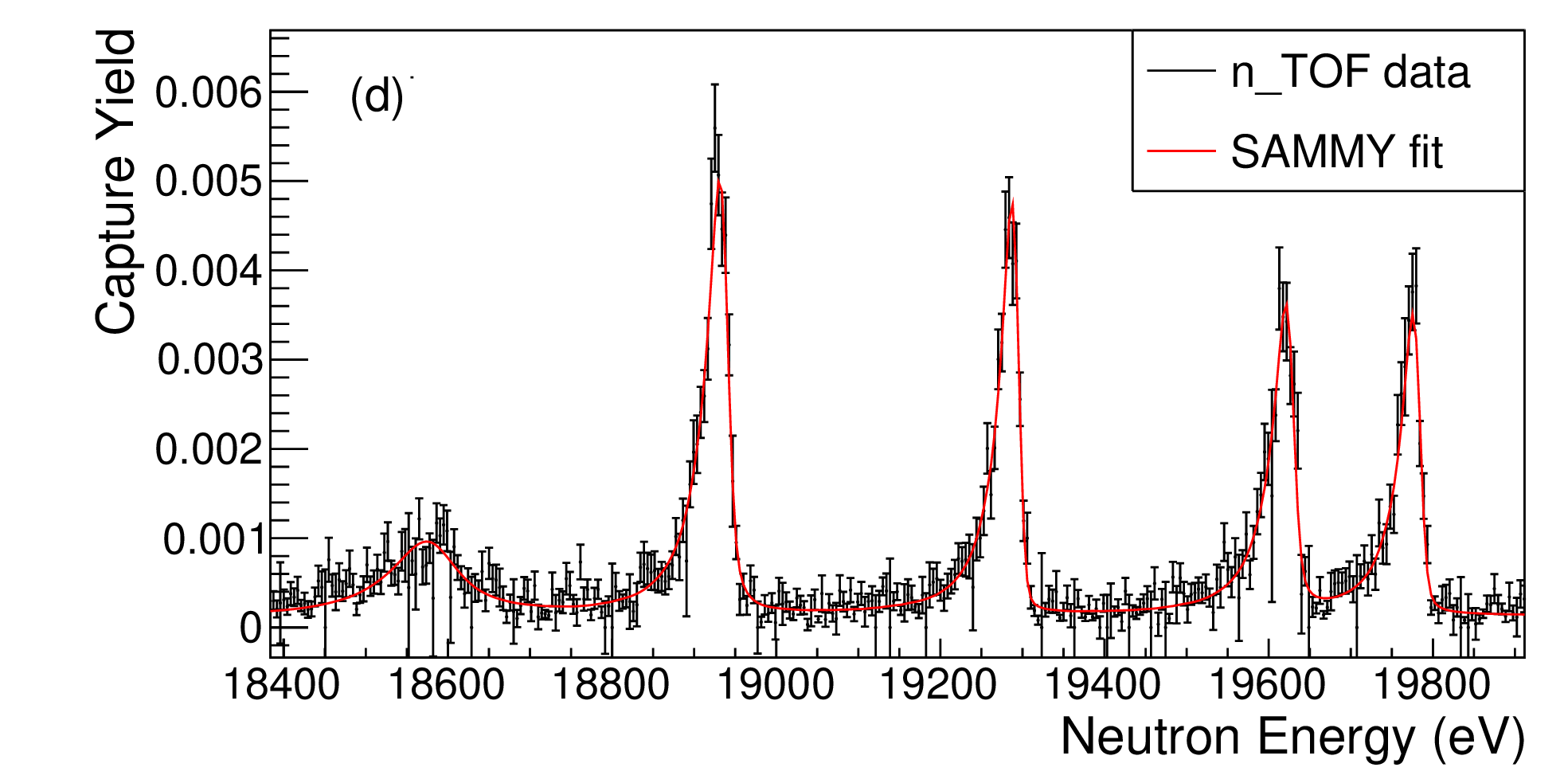}
\caption{(Color online) (a-d) Examples for some SAMMY fits of the experimental capture yield. \label{resonanceplot}}
\end{figure}

Neutron resonances in the capture yield were fitted with the multi-level, multi-channel R-matrix code SAMMY \cite{sammy}. SAMMY takes into account all experimental effects, 
such as multiple interaction events (multiple scattering) and self shielding, and the broadening of resonances due to thermal motion (Doppler Broadening) and the resolution of the experimental setup. 
In addition, the full sample composition, including impurities was taken into account. \\
Capture data themselves do not usually allow a reliable determination of individual resonance parameters, such as resonance spin $J$ and partial neutron and radiative widths,  $\Gamma_n$ and $\Gamma_\gamma$. 
In general, only energy and capture kernel $k$, defined as 
\begin{equation}
 k=g\frac{\Gamma_n \Gamma_\gamma}{\Gamma_n+\Gamma_\gamma},
\end{equation}
can be obtained reliably. 
The statistical factor $g$ is given by 
\begin{equation}
g=\frac{(2J+1)}{(2s+1)(2I+1)}
\end{equation}
where $J$ is the resonance spin, the neutron spin $s=1/2$, and the ground state spin of the target nucleus  $I(^{70}$Ge$)=0^{+}$, hence in our case $g=(2J+1)/2$. \\ 
Resonance structures could be resolved up to neutron energies of 40~keV. However, from about 25~keV the analysis of individual resonance parameters became increasingly
difficult, due to the worsening of the experimental resolution combined with lower statistics and increasing natural resonance widths.  
For this reason, an averaged cross section was determined from 25~keV onwards (see Sec. \ref{URRsec}).  \\
Resonance energies and capture kernels $k$ are shown in Tables \ref{tablekernel} and \ref{tablekernel2} for neutron resonance energies below, and above 25~keV, respectively.  
Examples of resonance fits in various neutron energy ranges are shown in Fig. \ref{resonanceplot}. 
Uncertainties in Tables \ref{tablekernel} and \ref{tablekernel2} are fit uncertainties only.

Systematic uncertainties in the capture kernels are due to the PHWT (2\%), the normalisation (1\%), the neutron flux (1\% for $E_n<3$~keV, 3.5\% for $E_n>3$~keV), and the sample enrichment (1\%). 
This amounts to total systematic uncertainties of 2.7\% below, and 4.3\% above 3~keV neutron energy, respectively. In total, we fitted 110 resonances up to energies of 40~keV, of which 90 were 
not listed in previous evaluations.  \\

Average resonance parameters, namely the average radiative width $\overline{\Gamma}_\gamma$ and the average resonance spacing $D_0$, were determined using the resonances 
below 25~keV assuming there are no unresolved doublets.
There are 35 strong resonances for which the SAMMY fit yielded $\Gamma_n > 10 \times \Gamma_\gamma$. 
As $k \approx g \Gamma_\gamma$ for these resonances, we used 
the kernels determined from n\_TOF data and the spins from a transmission measurement \cite{HH80} to estimate 
the distribution of individual $\Gamma_\gamma$ values in terms of the average radiative width $\overline{\Gamma}_\gamma$ and 
the width of the distribution $\sigma_{\Gamma_\gamma}$. Using the same method as in Ref.~\cite{LB19}, 
namely the maximum likelihood fit assuming a gaussian distribution of $\Gamma_\gamma$ values, we obtained $\overline{\Gamma}_\gamma=205(12)$ meV 
and $\sigma_{\Gamma_\gamma}=70(10)$. Our value of the average radiative width is compatible with the literature values 185(30)~meV~\cite{Mughabghab} 
and 185(50)~meV~\cite{RIPL3}.

To determine $D_0$ we have adopted the neutron strength functions for $s$-, $p$- and $d$-wave neutrons $S_0=2.1\times 10^{-4}$, $S_1=1.5\times 10^{-4}$
and $S_2=3\times 10^{-4}$, and channel radius of 7 fm from Ref.~\cite{Mughabghab}. The numbers of observed resonances above certain thresholds in the 
resonance kernel were compared to the simulated resonance sequences assuming the validity of the statistical model as in Ref.~\cite{Lerendegui-Marco}. 
Our data are perfectly consistent with $D_0=830(100)$ eV, this value is compatible within 2$\sigma$ with the literature value of 1170(230)~eV~\cite{Mughabghab,RIPL3}.

Present values of $D_0$ and $\overline{\Gamma}_\gamma$ in combination with adopted $S_0$ and $S_1$ imply that the neutron width of the strongest $p$-wave 
resonances can reach the values of the strongest $s$-wave resonances already at $E_n$ of a few keV. The determination of $S_0$ and $S_1$ without firm $\ell$ 
assignment is thus impossible, we can merely state that the literature values are reasonable. For example, simulations with a half value of $S_1$ during the 
procedure of obtaining the $D_0$ yielded inconsistent $D_0$ values for different adopted thresholds.

\begin{table}[!htb]
\caption{\label{tablekernel} Resonance energies $E_R$ and kernels $k$ up to 25~keV determined with SAMMY. The uncertainties listed are from  the fitting procedure. }
\begin{ruledtabular}
\begin{tabular}{ccccc}
$	E_R$ (eV)		&	$	k $ (meV) 				&	&	$	E_R$ (eV)	&	$	k $ (meV) 				\\ \hline
$	152.38	\pm	0.01	$	&	$	0.058	\pm	0.002	$	&	&	$	12655.7	\pm	0.3	$	&	$	148.4	\pm	5.1	$\\
$	1118.4	\pm	0.1	$	&	$	150.5	\pm	5.7	$	&	&	$	13270.0	\pm	4.6	$	&	$	206	\pm	13	$\\
$	1474.23	\pm	0.01	$	&	$	140.9	\pm	0.9	$	&	&	$	13326.1	\pm	0.4	$	&	$	168.2	\pm	8.3	$\\
$	1484.17	\pm	0.02	$	&	$	11.5	\pm	0.2	$	&	&	$	13867.8	\pm	0.6	$	&	$	194	\pm	11	$\\
$	1514.23	\pm	0.02	$	&	$	24.0	\pm	0.2	$	&	&	$	13887.7	\pm	1.0	$	&	$	47.2	\pm	5.1	$\\
$	1953.16	\pm	0.02	$	&	$	29.8	\pm	0.4	$	&	&	$	14331.7	\pm	0.5	$	&	$	121.7	\pm	5.7	$\\
$	2358.51	\pm	0.06	$	&	$	7.6	\pm	0.2	$	&	&	$	14800.6	\pm	0.5	$	&	$	117.2	\pm	5.9	$\\
$	2652.83	\pm	0.07	$	&	$	9.8	\pm	0.3	$	&	&	$	15118.2	\pm	1.0	$	&	$	30.6	\pm	2.6	$\\
$	3170.27	\pm	0.04	$	&	$	46.6	\pm	0.8	$	&	&	$	15705.2	\pm	0.8	$	&	$	223.8	\pm	8.9	$\\
$	3224.02	\pm	0.05	$	&	$	24.2	\pm	0.5	$	&	&	$	16005.8	\pm	0.4	$	&	$	326	\pm	10	$\\
$	3846.2	\pm	0.1	$	&	$	11.5	\pm	0.4	$	&	&	$	16366.6	\pm	0.6	$	&	$	222	\pm	15	$\\
$	3853.7	\pm	0.3	$	&	$	3.5	\pm	0.3	$	&	&	$	16402.7	\pm	0.8	$	&	$	136.5	\pm	7.9	$\\
$	4290.40	\pm	0.05	$	&	$	60.3	\pm	1.0	$	&	&	$	16900.8	\pm	0.6	$	&	$	183.9	\pm	8.8	$ \\
$	4397.5	\pm	0.1	$	&	$	150.5	\pm	2.1	$	&	&	$	17032.8	\pm	0.7	$	&	$	268	\pm	17	$ \\
$	5157.0	\pm	0.3	$	&	$	5.8	\pm	0.4	$	&	&	$	17358.1	\pm	0.4	$	&	$	225	\pm	12	$	\\
$	5530.9	\pm	0.1	$	&	$	45.5	\pm	1.2	$	&	&	$	17706.4	\pm	1.0	$	&	$	45.5	\pm	3.6	$	\\
$	5602.0	\pm	0.5	$	&	$	271.0	\pm	4.2	$	&	&	$	17937.0	\pm	1.1	$	&	$	41.1	\pm	3.9	$	\\
$	6035.6	\pm	0.1	$	&	$	45.6	\pm	2.5	$	&	&	$	18614.7	\pm	3.1	$	&	$	186.3	\pm	9.7	$	\\
$	6590.8	\pm	0.3	$	&	$	17.2	\pm	0.9	$	&	&	$	18963.9	\pm	0.6	$	&	$	362	\pm	28	$	\\
$	6796.8	\pm	0.3	$	&	$	232.4	\pm	4.0	$	&	&	$	19319.4	\pm	0.5	$	&	$	278	\pm	12	$	\\
$	7259.7	\pm	0.1	$	&	$	120.5	\pm	4.8	$	&	&	$	19654.5	\pm	0.8	$	&	$	254	\pm	28	$	\\
$	7591.0	\pm	0.2	$	&	$	99.8	\pm	4.0	$	&	&	$	19809.1	\pm	0.6	$	&	$	209	\pm	11	$	\\
$	7669.2	\pm	0.2	$	&	$	36.8	\pm	1.5	$	&	&	$	20123.7	\pm	1.4	$	&	$	59.3	\pm	5.2	$	\\
$	8289.2	\pm	0.1	$	&	$	154.4	\pm	5.2	$	&	&	$	20272.3	\pm	0.7	$	&	$	278	\pm	14	$	\\
$	8663.9	\pm	0.6	$	&	$	41.2	\pm	3.3	$	&	&	$	20714.5	\pm	0.9	$	&	$	159.7	\pm	7.8	$	\\
$	8699.4	\pm	1.2	$	&	$	208.8	\pm	6.8	$	&	&	$	20883.5	\pm	1.1	$	&	$	88.6	\pm	6.4	$	\\
$	8723.5	\pm	0.4	$	&	$	67.5	\pm	4.2	$	&	&	$	21669.4	\pm	5.1	$	&	$	117	\pm	22	$	\\
$	9395.9	\pm	0.2	$	&	$	148.5	\pm	6.7	$	&	&	$	21697.9	\pm	0.9	$	&	$	127	\pm	58	$	\\
$	9957.0	\pm	1.4	$	&	$	241.5	\pm	6.9	$	&	&	$	22276.3	\pm	0.8	$	&	$	267	\pm	26	$	\\
$	10018.5	\pm	0.4	$	&	$	56.0	\pm	3.3	$	&	&	$	22734.8	\pm	2.4	$	&	$	188	\pm	12	$	\\
$	10118.2	\pm	0.4	$	&	$	42.7	\pm	2.2	$	&	&	$	23128.7	\pm	1.0	$	&	$	190	\pm	12	$	\\
$	10367.6	\pm	1.2	$	&	$	242.8	\pm	7.6	$	&	&	$	23556.7	\pm	1.6	$	&	$	53.9	\pm	6.3	$	\\
$	10505.0	\pm	0.2	$	&	$	237.4	\pm	7.7	$	&	&	$	23775.1	\pm	1.5	$	&	$	150.1	\pm	9.8	$	\\
$	11648.7	\pm	0.6	$	&	$	28.4	\pm	2.1	$	&	&	$	23916.7	\pm	0.8	$	&	$	440	\pm	32	$	\\
$	11838.5	\pm	0.3	$	&	$	232.0	\pm	6.9	$	&	&	$	24065.4	\pm	3.9	$	&	$	167	\pm	13	$	\\
$	12310.0	\pm	0.5	$	&	$	46.4	\pm	2.9	$	&	&	$	24685.1	\pm	0.9	$	&	$	329	\pm	15	$	\\
$	12399.2	\pm	0.2	$	&	$	253.0	\pm	8.6	$	&	&	$				$	&	$				$	\\
\end{tabular}

\end{ruledtabular}
\end{table}

\begin{table}[!htb]
\caption{\label{tablekernel2} Resonance energies $E_R$ and kernels $k$ from 25 keV determined with SAMMY.
The uncertainties listed are from  the fitting procedure. Some of these resonances may be more complex structures, which could not be separated due to the worsening of the
experimental resolution, and the increasing natural resonance widths. }
\begin{ruledtabular}
\begin{tabular}{ccccc}
$	E_R$ (eV)		&	$	k $ (meV) 				&	&	$	E_R$ (eV)	&	$	k $ (meV) 				\\ \hline
$	25011.8	\pm	2.1	$	&	$	101.5	\pm	8.3	$	&	&	$	32543.6	\pm	2.4	$	&	$	125	\pm	12	$	\\
$	25348.5	\pm	1.0	$	&	$	293	\pm	16	$	&	&	$	32768.1	\pm	3.8	$	&	$	94	\pm	12	$	\\
$	25656.7	\pm	1.5	$	&	$	94.2	\pm	8.9	$	&	&	$	34382.1	\pm	3.1	$	&	$	212	\pm	23	$	\\
$	25857.2	\pm	2.2	$	&	$	111	\pm	10	$	&	&	$	34681.5	\pm	7.3	$	&	$	121	\pm	21	$	\\
$	26039.4	\pm	6.6	$	&	$	272	\pm	19	$	&	&	$	35260	\pm	16	$	&	$	121	\pm	30	$	\\
$	26670.5	\pm	1.1	$	&	$	358	\pm	16	$	&	&	$	35605.1	\pm	7.1	$	&	$	456	\pm	49	$	\\
$	26925.8	\pm	2.2	$	&	$	115	\pm	10	$	&	&	$	35939.2	\pm	5.9	$	&	$	88	\pm	21	$	\\
$	27264.4	\pm	1.0	$	&	$	554	\pm	56	$	&	&	$	36198.9	\pm	3.1	$	&	$	423	\pm	49	$	\\
$	27706.5	\pm	1.1	$	&	$	320	\pm	17	$	&	&	$	36266.2	\pm	2.7	$	&	$	330	\pm	40	$	\\
$	27892.9	\pm	2.1	$	&	$	402	\pm	21	$	&	&	$	36585	\pm	15	$	&	$	219	\pm	35	$	\\
$	28137.4	\pm	2.0	$	&	$	254	\pm	15	$	&	&	$	37033.5	\pm	7.0	$	&	$	207	\pm	30	$	\\
$	29161.1	\pm	2.4	$	&	$	493	\pm	32	$	&	&	$	37185.2	\pm	3.8	$	&	$	343	\pm	33	$	\\
$	30035.2	\pm	1.4	$	&	$	381	\pm	19	$	&	&	$	37517.4	\pm	3.0	$	&	$	347	\pm	50	$	\\
$	30294.4	\pm	2.3	$	&	$	181	\pm	12	$	&	&	$	37864.3	\pm	3.1	$	&	$	208	\pm	24	$	\\
$	30666.4	\pm	1.3	$	&	$	360	\pm	16	$	&	&	$	38332.3	\pm	8.5	$	&	$	146	\pm	25	$	\\
$	31629.6	\pm	2.4	$	&	$	116	\pm	13	$	&	&	$	38507.7	\pm	3.7	$	&	$	177	\pm	18	$	\\
$	31845.7	\pm	1.4	$	&	$	388	\pm	27	$	&	&	$	39004.6	\pm	6.2	$	&	$	464	\pm	200	$	\\
$	32047.7	\pm	2.4	$	&	$	208	\pm	18	$	&	&	$	39871.6	\pm	5.6	$	&	$	315	\pm	39	$	\\
$	32405.2	\pm	6.6	$	&	$	172	\pm	22	$	&	&						&						\\
\end{tabular}

\end{ruledtabular}
\end{table}

\subsection{Unresolved Resonance Region \label{URRsec}}
\begin{figure}[!htb]
\includegraphics[width=7.5 cm]{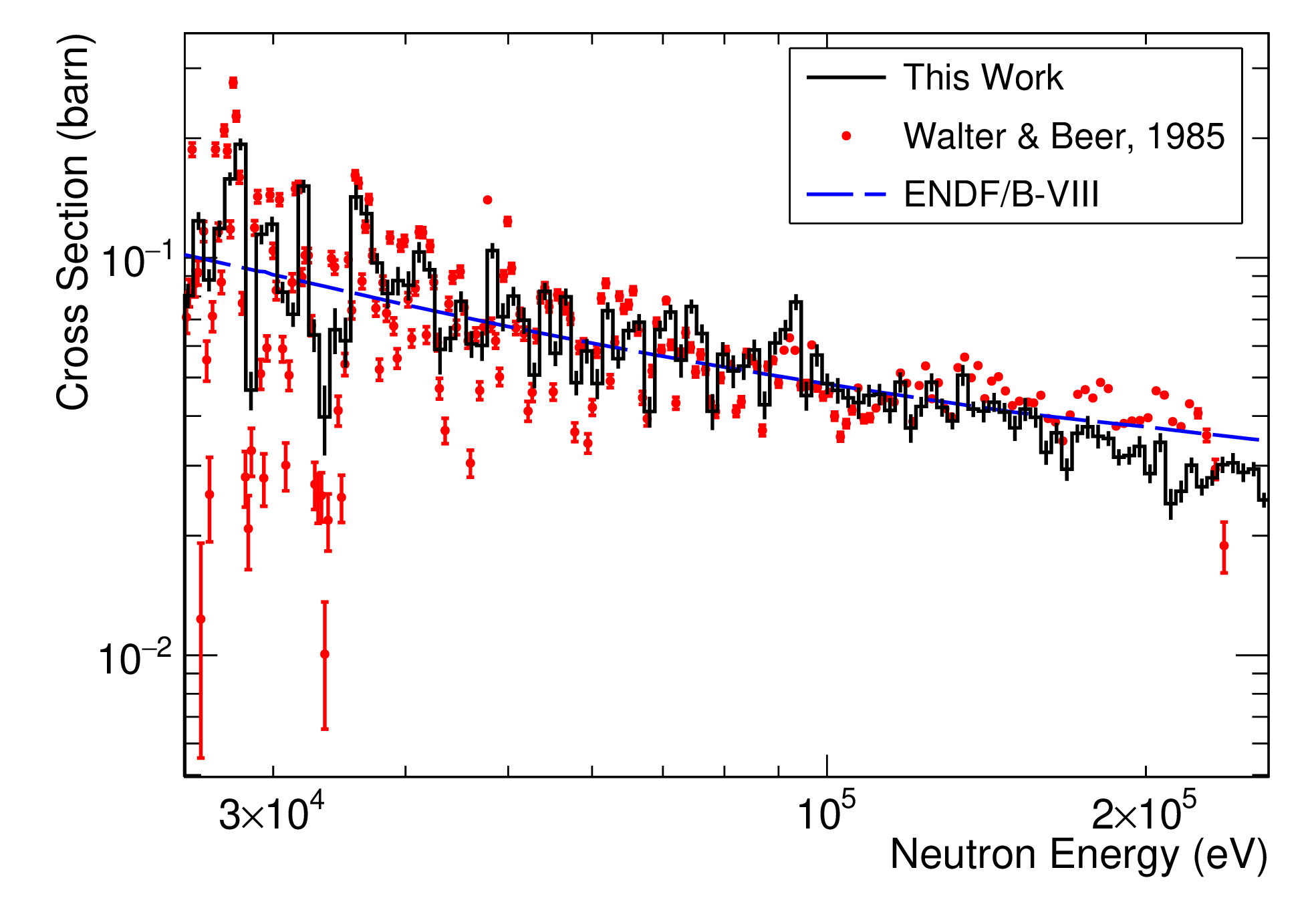}
\caption{(Color online) Neutron capture cross section with statistical uncertainties in the unresolved resonance region from 25 to 300 keV. The data obtained in this work are compared to experimental results by Walter and Beer \cite{WB85} and the 
ENDF/B-VIII evaluation \cite{ENDF8}.     \label{xs}}
\end{figure}

The threshold for observing individual resonances increases with energy, and from about 25~keV a non-negligible fraction of smaller resonances is missed, 
which would result in an underestimation of the cross section. Therefore, we determined an averaged neutron capture cross section between 25 keV and 300 keV from the neutron capture yield. 
The capture yield was corrected for multiple scattering and self shielding effects using a Monte Carlo code which takes into account the sample composition and geometry, as well as neutron scattering
and capture cross sections. These corrections to the yield amounted to 6-7\%. A conservative estimate of 20\% uncertainty in these corrections results in a 1.4\% uncertainty of the corrected 
neutron capture yield. \\
The unresolved cross section was obtained after subtraction of backgrounds due to neutron scattering (see Sec. \ref{bgsub}), and contributions due to impurities. In the present case, the
main impurity in the sample is $^{72}$Ge with a content of 2.23\%. Other germanium isotopes, as well as oxygen in the sample have negligible contributions to the measured capture yield,
either because of their low abundance in the sample, or because of their small cross sections.
We used ENDF/B-VIII \cite{ENDF8} cross sections to estimate the contribution of the $^{72}$Ge impurity to the measured cross section yielding a 
background at a level of 1-2\%.  At present there is no experimental information on the $^{72}$Ge($n,\gamma$) cross section in the relevant energy range
and the cross section recommended by the ENDF/B-VIII evaluation 
is based on average resonance parameters obtained at lower neutron energies by Maletski et al. \cite{Mal68}. Assuming a 20\% uncertainty in the $^{72}$Ge cross section, the contribution to the uncertainty in the 
$^{70}$Ge($n,\gamma$) cross section is at most 0.4\%. 
Total systematic uncertainties of the unresolved cross section are 6\%,  consisting of 4.3\% systematic uncertainty as outlined in Sec. \ref{RRRsec}, 
the background subtraction with neutron filters (4\%), correction for sample impurities (0.4\%), and multiple scattering and self-shielding corrections (1.4\%). 
Figure \ref{xs} shows a plot of the unresolved cross section obtained in this work, compared to experimental results by Walter and Beer \cite{WB85}, and evaluated cross sections
published in the ENDF/B-VIII library. The cross sections determined in this work are in fair agreement with Walter and Beer, as well as ENDF/B-VIII over most of the neutron energy range covered, while being systematically smaller than ENDF/B-VIII and Walter and Beer above 150~keV.

\section{Stellar Cross Sections and Astrophysical Implications}
Maxwellian Averaged Cross Sections were determined by combining the cross section obtained from resonance parameters up to 25~keV neutron energy (resolved resonance region RRR), 
with the unresolved averaged cross section for neutron energies between 25 and 300~keV (unresolved resonance region URR). From 300~keV, we used cross sections recommended by the 
ENDF/B-VIII evaluation, but scaled by a factor 0.8 to take into account that the new experimental data indicate an overestimation of the cross section at higher neutron energies of 10-35\%. MACS values were calculated for $kT$ values between 5 and 100~keV using the formula
\begin{equation}
 \text{MACS} = \frac{2}{\sqrt{\pi}} \frac{1}{(kT)^{2}}\cdot \int_0^\infty E \sigma(E) \cdot \exp{\left(-\frac{E}{kT}\right)} \mbox{d}E
\end{equation}

\begin{table}[htb]
\caption{Maxwellian Averaged Cross Sections obtained from resonance data below, and  averaged cross sections above 25~keV neutron energy.
The uncertainty is the total uncertainty, including systematic and statistical uncertainties. \label{tablemacs}}
\begin{ruledtabular}
\begin{tabular}{cccc}
$kT$ (keV) &  \multicolumn{3}{c}{MACS (mb)} \\
                   & This work & Kadonis-1.0 \cite{kad03} & Walter and Beer \cite{WB85} \\ \hline
$	5	$ & $	212.4	\pm	9.3	$  & $	207.3	$  & $				$ \\
$	10	$ & $	159.7	\pm	7.1	$  & $	154.8	$  & $				$ \\
$	20	$ & $	115.8	\pm	5.6	$  & $	109.8	$  & $	112	\pm	6	$ \\
$	30	$ & $	94.2	\pm	4.9	$  & $	89.1\pm5.0	$  & $	92	\pm	5	$ \\
$	40	$ & $	80.8	\pm	4.3	$  & $	77.1	$  & $	81	\pm	5	$ \\
$	50	$ & $	71.5	\pm	3.9	$  & $	69.3	$  & $	75	\pm	4	$ \\
$	60	$ & $	64.6	\pm	3.6	$  & $	63.7	$  & $				$ \\
$	70	$ & $	59.3	\pm	3.3	$  & $		$  & $				$ \\
$	80	$ & $	55.1	\pm	3.0	$  & $	56.2	$  & $				$ \\
$	90	$ & $	51.7	\pm	2.9	$  & $		$  & $				$ \\
$	100	$ & $	48.9	\pm	2.7	$  & $	51.4	$  & $				$ \\

\end{tabular}

\end{ruledtabular}
\end{table}

\begin{figure}[!htb]
\includegraphics[width=9.0 cm]{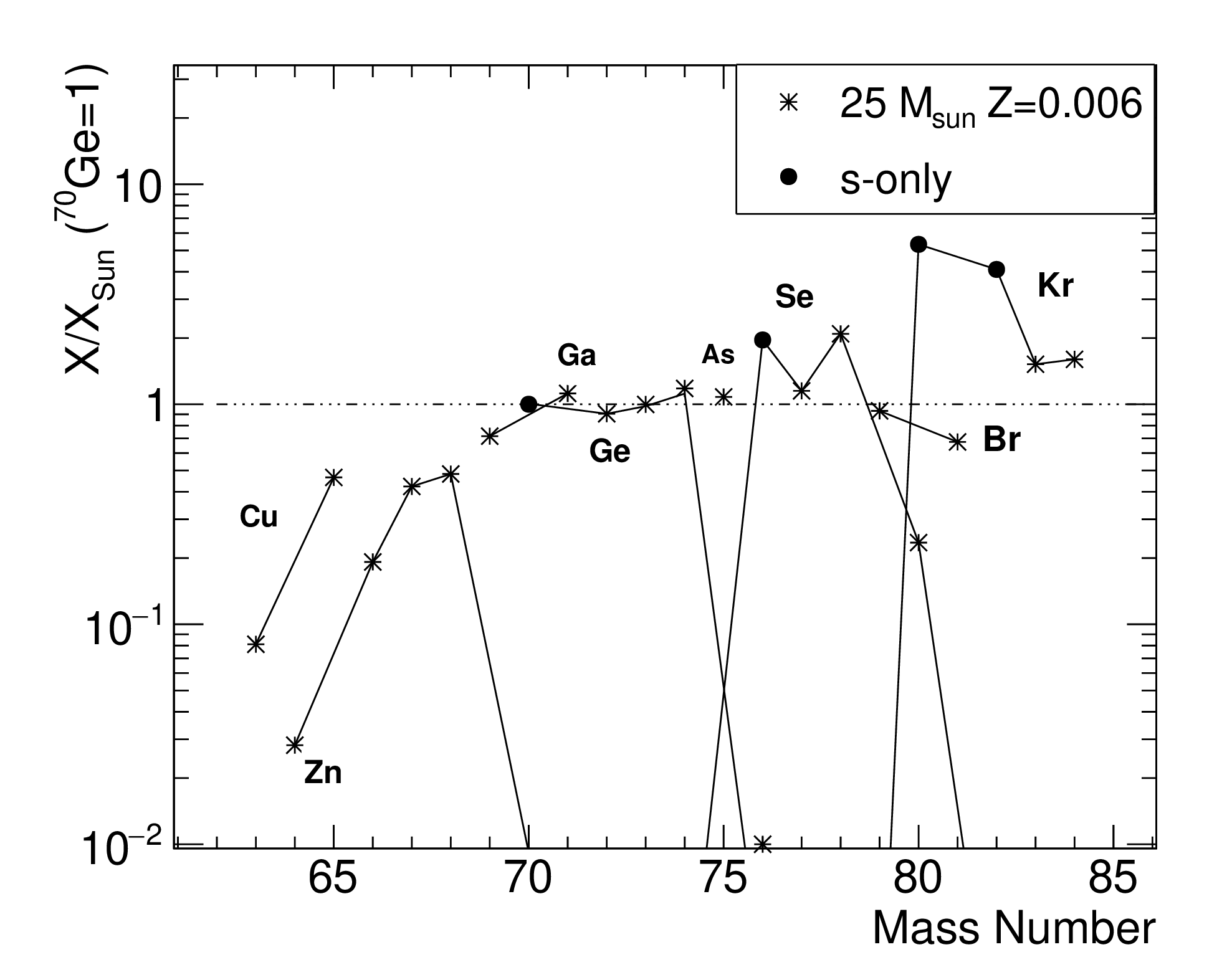}
\includegraphics[width=9.0 cm]{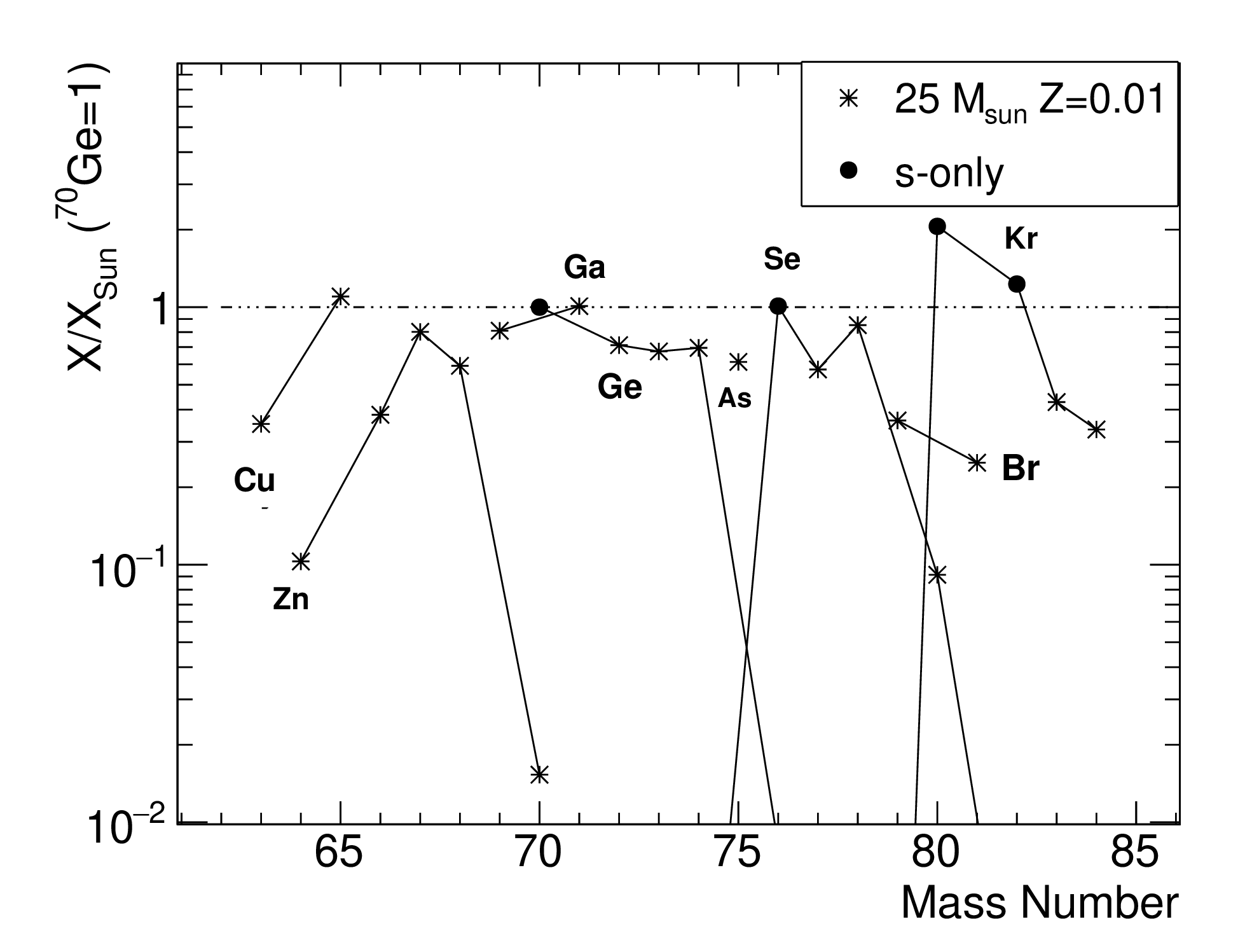}
\caption{Abundances produced in a 25 $M_{\odot}$ star for two initial metallicities $z$, normalised to solar system abundances for which contributions
from the main $s$ process and $p$ process have already been subtracted.  The distribution has been normalised to $^{70}$Ge. $s$-only isotopes are shown as full circles and isotopes of the same element are connected by solid lines. The $z=0.006$ metallicity model produces germanium isotopic abundances close to solar proportions, while the $z=0.01$ model provides a better global fit for A=60-80.    \label{macsompli}}
\end{figure}

Uncertainties taking into account all three components were determined in the following way: The common (absolute) uncertainty for RRR + URR MACSs $\sigma_{\text{RRR+URR}}$ is due 
to the weighting procedure (2\%), the sample enrichment (1\%), the neutron flux (3.5\%), normalisation (1\%), and counting statistics including uncertainties of the resonance 
fits (1\%). An additional uncertainty just applicable to the URR, $\sigma_{\text{URR}}$, comes from the background subtraction using filters (4\%), self shielding and multiple scattering
(1.4\%), and subtraction of the $^{72}$Ge background (0.4\%). We have assigned a 20\% uncertainty to the scaled ENDF/B-VIII cross sections ($\sigma_{\text{ENDF}}$) that were used for $E_n>300$ keV. 
The contribution of cross sections above 300~keV to the MACS is negligible for $kT$ values below 50~keV,
and between 2\% and 11\% for $kT=60-100$~keV. 
The total uncertainty of the MACS was then determined as
\begin{equation}
\sigma_{\text{tot}}=\sqrt{\sigma_{\text{RRR+URR}}^{2} + \sigma_{\text{URR}}^{2} + \sigma_{\text{ENDF}}^{2}}
\end{equation}
Hence, total uncertainties for MACSs values vary from 4.4\% at $kT=5$~keV to 5.6\% at $kT=100$~keV. \\
Table \ref{tablemacs} lists MACS values determined in this work from $kT=5$ to 100 keV. MACSs in this work are compared to recommended values in the Karlsruhe Astrophysical Database of Nucleosynthesis in Stars (KADoNiS) version 1.0 \cite{kad03},
which is widely used as reference for reaction rates in astrophysical calculations, and experimental MACSs determined by Walter and Beer \cite{WB85}. 
Agreement with Walter and Beer values is very good, and there is also good agreement with Kadonis-1.0, considering uncertainties. The trend of the MACS values with $kT$ is flatter for Walter and Beer, which can be explained by their higher averaged cross section at high neutron energies.\\
Because $^{70}$Ge is an even-even nucleus, the contribution of reactions on
excited target states in the stellar plasma is negligible up to plasma
temperatures of about 1.5 GK, which is well above the s-process
temperature range \cite{rau12a,rau12b}. Therefore the experimentally
determined MACS can directly be used as stellar MACS in astrophysical
calculations for the s process.\\
We have calculated $s$ process nucleosynthesis abundances for a massive  
25 $M_{\odot}$ star for two initial metallicities, below solar ($z=0.006$), and close to solar metallicity ($z=0.01$)\footnote{$z$ is defined as the mass fraction of elements heavier than helium, solar metallicity is $z_{\odot}=0.014$}, using the  multi-zone post-processing code {\sc mppnp} \cite{HDF08}. The neutron capture network, which is largely based on rates recommended in the latest released version of Kadonis, Kadonis-v0.3 \cite{kad03}, included  the new $^{70}$Ge($n,\gamma$)  MACSs obtained in this work, and recent results from $^{73}$Ge($n,\gamma$) \cite{LB19}.  Due to the very good agreement with previous results, there are only small changes in the resultant $^{70}$Ge abundances produced in massive stars (about 3\%) using the new $^{70}$Ge MACS. However, our data provide an independent confirmation of the only previous measurement at stellar energies, and in addition, include also MACS for $kT>50$~keV, thus improving the accuracy of stellar model predictions. \\
Figure \ref{macsompli} shows the $weak$ $s$ process abundances produced before supernova explosion, scaled to solar system abundances and normalised to $^{70}$Ge. Isotopes of the same element have been connected by solid lines. 
Contributions to the solar system abundances due to the $main$ $s$ process and the $p$ process have been subtracted using results by  Arlandini et al. \cite{AKW99} and Travaglio et al. \cite{Tra11}, respectively. The absolute contributions of these two nucleosynthesis processes have been determined by normalising to the $main$ $s$-only isotope $^{150}$Sm, and $p$-only isotope $^{84}$Sr, respectively. Hence, the solar system abundances shown in Fig. \ref{macsompli}  only contain contributions from the $weak$ $s$-process, and explosive nucleosynthesis processes ($r$ process,  alpha-rich freeze out). 
The results show that the $z=0.006$ model reproduces best the solar isotopic abundance pattern of germanium isotopes, while the model close to solar metallicity ($z=0.01$) provides a better global fit to the other $s$-only isotopes $^{76}$Se, $^{80}$Kr and $^{82}$Kr. Furthermore, the $z=0.01$ model indicates that the $weak$ $s$~process contributes a large fraction to solar $^{65}$Cu, $^{67,68}$Zn,  $^{69,71}$Ga, $^{72,73,74}$Ge, $^{75}$As  and $^{77,78}$Se abundances. For a firm conclusion about the absolute contribution of the $weak$ $s$-process however, stellar nucleosynthesis calculations using the new cross sections need to be implemented into calculations of Galactic Chemical Evolution, taking into account stars of different initial metallicities and masses, as well as contributions from explosive nucleosynthesis processes during the core collapse supernova explosion. 

\section{Summary}
We have measured $^{70}$Ge($n,\gamma$) cross sections up to 300 keV neutron energy at the neutron time-of-flight facility n\_TOF at CERN. Resonance capture kernels of 110 resonances were determined up to 40 keV, 
while averaged cross sections have been determined up to 300 keV. 
Maxwellian Averaged Cross Sections have been calculated, which are in very good agreement with the only other previous measurement of this 
reaction at stellar energies \cite{WB85}, providing an important independent confirmation
of stellar cross sections used in astrophysical calculations. The new MACSs combined with recent results for MACSs on  $^{73}$Ge($n,\gamma$) \cite{LB19} have been used to calculate abundances produced in two massive star models with sub-solar, and close to solar metallicity. Abundances for $s$-only isotopes match reasonably well with solar system values, in particular for the model close to solar metallicity, while the $=0.006$ model reproduces the germanium isotopic abundance pattern in our solar system.

  \section*{Acknowledgements}
This work was supported by the Austrian Science Fund FWF (J3503), the Adolf Messer Foundation (Germany), 
the UK Science and Facilities Council (ST/M006085/1), and the European Research Council ERC-2015-StG Nr. 677497. 
We also acknowledge support of Narodowe Centrum Nauki (NCN) under the grant (UMO-2016/22/M/ST2/00183) and from MSMT of the Czech Republic and the Croatian Science Foundation under the project IP-2018-01-8570.

\end{document}